\documentclass[12pt]{article}
\usepackage{amsmath}%
\usepackage{amsfonts}%
\usepackage{amssymb}%
\usepackage{graphicx}

\begin{document}

\title{Simulation study of traffic accidents in bidirectional traffic models}
\author{Najem Moussa \thanks{e-mail: najemmoussa@yahoo.fr}
\\ D\'{e}partement de Math\'{e}matique et Informatique,
\\ Facult\'{e} des Sciences, B.P. 20 - 24000 - El Jadida, Morocco}
 \maketitle

\begin{abstract}
Conditions for the occurrence of bidirectional collisions are
developed based on the Simon-Gutowitz bidirectional traffic model.
Three types of dangerous situations can occur in this model. We
analyze those corresponding to head-on collision, rear-end
collision and lane-changing collision. Using Monte Carlo
simulations, we compute the probability of the occurrence of these
collisions for different values of the oncoming cars density. It
is found that the risk of collisions is important when the density
of cars in one lane is small and that of the other lane is high
enough. The influence of different proportions of heavy vehicles
is also studied. We found that heavy vehicles cause an important
reduction of traffic flow on the home lane and provoke an increase
of the risk of car accident.

Pacs numbers: 89.40.+k; 02.60.Cb; 45.70.Vn; 89.75.Fb

\end{abstract}

\section{Introduction}

In the past few years, the investigation of traffic flow using
cellular automata (CA) has become quite popular
\cite{chowd1,maer}. The most popular CA model for traffic flow on
one-lane roadway is the Nagel-Schreckenberg (NS) model \cite{ns}.
Despite its simplicity, the model is capable of capturing some
essential features observed in realistic traffic like density
waves or spontaneous formation of traffic jams. After that,
various generalizations and extensions of this one-lane model have
been considered and different other congested traffic states may
occur such as synchronous phase, wide moving jams and stop-and-go
phase \cite{helb,naga}.
\newline\ The NS model rule set was extended to a multi-lane
model and has been applied to the investigation of traffic flow in
highways [6-9]. A CA model of bidirectional vehicular traffic by
Simon and Gutowitz (SG) extends the case to vehicles moving in
opposite directions on a two-lane road where passing is allowed
\cite{simon}. When a driver encounters a slower forward moving
vehicle, a pass will be attempted. To do this, driver checks the
density of vehicles in front that have to be passed, i.e. the
local density. If this density is low enough the pass will be
performed on condition of checking the safety criteria on the
oncoming lane. If the density of cars in one lane is small and
that of the other lane is high enough then wide jams occur in both
lanes. The occurrence of these wide jams are due principally to
the fact that almost all passing cars get stuck with oncoming cars
before returning to their home lanes. Recently, the SG model for
bidirectional traffic flow is revisited where some rules are
changed in order to prevent the occurrence of wide moving jams
\cite{mou2}.
\newline\ In modern society, traffic accidents and congestion continue
to plague the transportation managers. Accident analysis and
prevention is one of the most important aspects of interest in
traffic management because it is associated with human life.
Recently, cellular automata models have been used to investigate
the probability of the occurrence of car accidents. Boccara et al.
\cite{bocca} have been the first authors to propose conditions for
car accidents to occur in the deterministic NS model. The first
condition is that the number of empty cells in front of the car
(gap) is smaller than the speed limit, the second condition is
that the car ahead is moving, and the last condition is that the
moving car ahead is suddenly stopped at the next time step. Using
these conditions, analytic results of the probability of car
accidents are obtained in some special cases of the NS model with
periodic boundary conditions \cite{huang1,huang2}. General
numerical results concerning the probability of car accidents are
reported for the nondeterministic NS model [15-17]. The
probability for an accident to occur is found to be proportional
to the product of the fraction of stopped cars and the traffic
flow \cite{yang2}. Conditions for the occurrence of car accidents
in the NS model are developed based on the driver reaction time
\cite{mou3}. The probability for car accidents to occur has been
extensively investigated in the velocity effect model
\cite{jiang2}, two-lane CA model \cite{mou4}, NS model with open
boundaries \cite{yang3}, NS model with speed limit zone
\cite{zhang}, NS model with velocity-dependent randomization
\cite{yang4} and NS model with quenched randomness
\cite{yang5,yang6}
\newline\ The purpose of the present paper is precisely the
investigation of the probability of collisions in bidirectional
traffic models and to analyze the effect of heavy vehicles on the
traffic flow and on the probability of car accidents. The paper is
organized as follows. In Sec. $2$, we describe the model for
bidirectional traffic. Here, we consider the SG bidirectional
model revisited \cite{mou2}. In Sec. $3$, we introduce conditions
for the occurrence of car accidents. In Sec. $4$, we present our
simulation results and discussions. Finally, we conclude with a
summary in Sec. $5$.
\section{Model}
The SG model is a probabilistic CA which consists of vehicles
moving on two opposite lanes of $L$ cells with periodic boundary
conditions. Each cell is either empty, or occupied by just one
vehicle. In our model, there exist two types of lanes:
lane$^{[+]}$ and lane$^{[-]}$. On lane$^{[+]}$, two kinds of
vehicles move from left to right: cars and heavy vehicles. We
suppose that passing is forbidden for heavy vehicles and only the
cars can change lanes. In the model, we suppose that no heavy
vehicles exist on lane$^{[-]}$ and only cars move from right to
left. We denote by vehicle$^{[+]}$ the vehicle moving on the
lane$^{[+]}$ with positive direction and vehicle$^{[-]}$ the
vehicle moving on the lane$^{[-]}$ with negative direction (see
figure 1). We denote by $x$ and $v$ the position and the speed of
a vehicle at time $t$ respectively. To distinguish between cars
and heavy vehicles, we suppose that the maximal speed of the heavy
vehicles is lower to the one of the cars. Hence, we set
$v_{max}=3$ for heavy vehicles and $v_{max}=5$ for cars. To
distinguish between different interacting vehicles, several gaps
and speeds are introduced. $gap_{same}$ ($gap_{opp}$): the number
of unoccupied sites in front of a vehicle on the same (opposite)
lane. $gap_{behind}$: the number of unoccupied sites behind the
vehicle, on the opposite lane. $v_{same}$ ($v_{opp}$): the speed
of the vehicle ahead on the same (opposite) lane. On the aim of
making more compact the rules of the bidirectional model, several
logical functions are introduced. \textbf{H}: true if the vehicle
is on its home lane. \textbf{T}: true if the vehicle is a heavy
vehicle. \textbf{oncoming}: true if vehicle$^{[+]}$ and
vehicle$^{[-]}$ are face-to-face on the same lane.
\textbf{Space1}: true if ($gap_{same} < l_{pass}$) AND
($gap_{opp}>l_{security}$) AND ($gap_{behind} > l_{back}$).
\textbf{Space2}: true if $gap_{behind} > l_{back}$. The parameters
$l_{pass}$, $l_{back}$ and $l_{security}$ are defined by the
following. $l_{pass}$: if $gap_{same} < l_{pass}$ AND \textbf{H}
then a pass may be attempted. $l_{back}$: the distance a driver
looks back for obstacles on the passing lane. $l_{security}$: if
$gap_{same} < l_{security}$ AND not(\textbf{H}) then the vehicle
returns immediately to its home lane. $D_{L}$: local density: the
fraction of the $l_{density} = 2v_{max}+1 sites$ in front of the
given vehicle which are occupied; $D_{limit} = 1/l_{density}$: the
maximum local density for a safe pass.
\newline\ At each discrete time-step
$t\rightarrow t+1$ the system update is performed in parallel for
all cars according to the following subrules :
\newline\ \textbf{i.} \textbf{Lane changing rules}:
\newline\ 1. IF (not(\textbf{T}) AND \textbf{H} AND \textbf{Space1}
AND ($D_{L}\leq D_{limit}$) AND ($rand < p_{change}$)) THEN change
lane
\newline\ 2. IF (not(\textbf{T}) AND not(\textbf{H}) AND
(($gap_{same} < l_{security}$) OR \textbf{Space2})) THEN change
lane.
\newline\ The first condition concerns cars on their home lane that
want to change lane. When a driver encounters a slower forward
moving vehicle, a pass is attempted. However, the pass will only
be initiated if there is room far enough ahead on the passing
lane, and the number of vehicles in front of the car it would like
to pass is small. Passing occurs randomly, even if all these
conditions are met; the probability of changing lanes is denoted
$p_{change}$. The second condition concerns cars in the midst of
passing. They return to their home lane if forced to by an
oncoming vehicle, or if there is space enough on the home lane
that they can return without braking.
\newline\ \textbf{ii.} \textbf{Forward moving rules}:
\newline\ 1. IF ($v \neq v_{max}$) THEN $v = v + 1$
\newline\ 2. IF((\textbf{oncoming}) AND ($gap_{same}\leq(2 × v_{max}-1))$)
THEN $v = gap_{same}/2$
\newline\ 3. IF ( (not(\textbf{oncoming})) AND ($v
> gap_{same})$) THEN $v = gap_{same}$
\newline\ 4. IF (\textbf{H} AND ($v\geq 1$) AND ($rand < p_{decel}$) AND
not(\textbf{oncoming})) THEN $v =v-1$
\newline\ 5. IF (\textbf{H} AND (\textbf{oncoming}) AND ($v
\geq 1)$ ) THEN $v = v-1$.
\newline\ If the vehicle is a vehicle$^{[+]}$ then the vehicle moved forward
according to: $x \leftarrow x+v$. But, if the vehicle is a
vehicle$^{[-]}$ then the vehicle moved backward according to $x
\leftarrow x-v$.
\newline\ The rule (1) reflects the tendency of drivers to drive as fast as
possible without exceeding the maximum speed limit. Rule (2)
rapidly decelerates the vehicle if there is an oncoming vehicle
too close. Rule (3) is intended to avoid collision between the
vehicles of the same type. Rule (4) randomly decelerates the
vehicle if it is on its home lane; if it is passing, it never
decelerates randomly. Finally, rule (5) breaks the symmetry
between the lanes, and thus prevents the emergence of a super jam,
i.e., a jam which may occur when each of an adjacent pair of
vehicle$^{[+]}$ and vehicle$^{[-]}$, one on each lane, tries to
pass simultaneously.
\section{Conditions for the occurrence of car accidents}
In order to make effective plans or strategies of comprehensive
traffic safety, it is important to evaluate objectively the risk
of traffic accidents. However, the empirical data of car accidents
cannot be established from real life traffic since the traffic
accident itself is a kind of rare event. In reality, no one can
make real crash tests with living drivers. Thus, theoretical
investigations using statistical physics are highly desirable to
evaluate the effects of traffic safety in terms of number of
traffic accident. Recently, cellular automata models have been
extended to investigate the probability of car accidents in
unidirectional traffic roads [12-26]. In this paper, we shall
investigate the probability of car accidents in bidirectional
traffic flow models. We distinguish three types of car accidents:
rear-end collision, lane-changing collision and head-on collision.
\newline\ $\mathbf{i.}$ Rear-end collision is between two vehicles evolving
in the same lane and with the same direction. Based on the delayed
reaction time of the successor vehicle and the unexpected abrupt
deceleration of the predecessor vehicle, the conditions for the
occurrence of rear-end collision are expressed \cite{mou3}. When
the vehicle ahead with speed $v^{[+]}_{k+1}\left( t\right)$ do an
abrupt deceleration, its velocity will be reduced to
$v^{[+]}_{k+1}\left( t+1\right)$ at time $t+1$. Thus, if the
covered distance during the delayed reaction time $\tau $ of the
successor vehicle is enough to reach the next time position of the
vehicle ahead, rear-end collision happens most likely on the road.
Hence, the conditions for the occurrence of rear-end collision
with respect to abrupt deceleration of the vehicle ahead are as
follows:
\begin {eqnarray}\label{eq:test}
\ \tau v^{[+]}_{k}\left(t\right) >gap^{k}_{same}(t)+v^{[+]}_{k+1}(t+1)\\
v^{[+]}_{k+1}\left(t\right)-v^{[+]}_{k+1}\left( t+1\right)\geq v_d
\end{eqnarray}
where $gap^{k}_{same}(t)$ is the number of unoccupied sites in
front of the vehicle (k) on the same lane at time t. The parameter
$v_d$ is the deceleration limit beyond which a risk of the
occurrence of rear-end collision exists.
\newline\ $\mathbf{ii.}$ Lane-changing collision is between a
vehicle$^{[+]}$ on lane$^{[-]}$ that want to return to its home
lane and its following vehicle$^{[+]}$ on lane$^{[+]}$. An
unexpected lane changing of a vehicle$^{[+]}$ may lead to a
collision on lane$^{[+]}$, if the following vehicle$^{[+]}$ cannot
decelerate enough \cite{mou4}. In terms of delayed reaction time
$\tau$ of the following vehicle, we can express the conditions for
the occurrence of lane-changing collision as:
\begin {eqnarray}\label{eq:test}
\ \textit{vehicle}^{[+]}_{(k+1)} \textit{changes lane at time t} \\
 \tau v^{[+]}_{k}\left( t\right)>gap^{k}_{same}(t)+v^{[+]}_{k+1}(t+1)
\end{eqnarray}
If the distance covered during the delayed reaction time $\tau $
of the following vehicle$^{[+]}$ (k) on lane$^{[+]}$ is enough to
reach the next time position of the vehicle$^{[+]}$ ahead that
changed lanes (k+1), lane-changing crash happens most likely on
lane$^{[+]}$.
\newline\ $\mathbf{iii.}$ Head-on collision is between a vehicle$^{[+]}$ and
a vehicle$^{[-]}$ that evolve on the same lane and opposites
directions. This type of collision leads usually to fatal
injuries. The conditions for the occurrence of head-on collision
are:
\begin {eqnarray}\label{eq:test}
\ \textit{vehicle}^{[+]} \textit{and vehicle}^{[-]} \textit{are face to face} \\
 \tau v^{[+]}_{k}\left( t\right)>gap^{k}_{same}(t)-v^{[-]}_{k+1}(t+1)
\end{eqnarray}
If the distance covered during the delayed reaction time $\tau $
of the vehicle$^{[+]}$ (k) is enough to reach the next time
position of the vehicle$^{[-]}$ (k+1) ahead, which moves in the
opposite direction, fatal car accident happens most likely.
\section{Numerical results}
We simulate the bidirectional traffic model described above on a
lattice of $L$ sites with closed boundary conditions. The density
$\rho^{[+]}$ (resp. $\rho^{[-]}$) is defined as
$\rho^{[+]}=N^{[+]}/L$ (resp. $\rho^{[-]}=N^{[-]}/L$), where
$N^{[+]}$ (resp. $N^{[-]}$) is the number of vehicles$^{[+]}$
(resp. vehicles$^{[-]}$). We denote by $N_{HV}^{[+]}$ the number
of heavy vehicles in the circuit. In this article, without losing
generality, the model parameters are assumed to be as follows:
$l_{pass} = v$, $l_{back} = v_{max}$, $l_{security} = 2×v_{max}
+1$, $p_{change} = 0.5$, and $p_{decel} =0.3$. The maximal
velocity of the cars is $v_{\max }=5$ and that of the heavy
vehicles is $v_{\max }=3$. The system size is given by $L=2000$.
The delayed reaction time is assumed to be $\tau =1s$ and the
deceleration limit is $v_d=2$.
\subsection{Traffic flow and lane-changing frequency}
Clearly, the traffic flow of vehicles$^{[+]}$, the lane-changing
frequency of cars$^{[+]}$ and the occurrence of dangerous
situations, all should depend on the traffic density on the
oncoming lane. As in reference \cite{mou2}, we shall analyze the
case of two different oncoming lane densities. So, we choose
$\rho^{[-]}=0.05$ and $\rho^{[-]}=0.30$, representing densities
below and above the critical density in the NS model,
respectively. As in unidirectional traffic models, when
$\rho^{[+]}$ is low, the presence of heavy vehicles in the circuit
leads to the formation of platoon of cars behind the heavy
vehicles. Yet, the presence of only one heavy vehicle can lead to
a drastic reduction of the average flow in the circuit. When
$\rho^{[-]}$ is small enough, the flow on the home lane is reduced
for low values of $\rho^{[+]}$. But, the flow on the oncoming lane
increases with increasing the fraction of heavy vehicles (Fig.
2a). Indeed, the presence of heavy vehicles in the home lane
encourages the pass of cars$^{[+]}$ on the oncoming lane. If
$\rho^{[-]}$ is relatively high, passing of cars$^{[+]}$ become
rare and the flow on the oncoming lane will be very small (Fig.
2b). Clearly, the heavy vehicles dominate the average flow on the
home lane as like as in one-lane systems.
\newline\ In figures 3a and 3b, we show the lane-changing frequency
of cars$^{[+]}$, as a function of $\rho^{[+]}$, for different
values of $\rho^{[-]}$. When $\rho^{[-]}$ is low, and in the
absence of heavy vehicles, the lane-changing rate of cars$^{[+]}$
is maximal at density $\rho^{[+]}$ around the critical density
separating the free-flow and congested regimes in the NS model
(Fig. 3a). If one heavy vehicle is present in the circuit, almost
all the cars$^{[+]}$ are organized into platoon behind the heavy
vehicle. Lane-changing rate of the cars$^{[+]}$ will be weakened,
compared to the situation where no heavy vehicle is present,
because of the reduction of their speeds and the higher value of
their local densities. Additional heavy vehicles may divide the
platoon into small ones and then decreases the local densities.
This may increase the lane-changing rate of cars$^{[+]}$ and then
their flow will increase on the oncoming lane. In the case where
$\rho^{[-]}$ is high enough, passing of cars$^{[+]}$ become rare
and the lane-changing frequency will be very small (Fig. 3b).
Moreover, we find that higher values of the lane-changing
frequency occur at very low values of $\rho^{[+]}$. Since free
available spaces may exist on lane$^{[+]}$, cars$^{[-]}$ can pass
other cars$^{[-]}$ and then impede traffic on lane$^{[+]}$. In
this situation, lane-changing maneuvers can be performed by the
fast cars$^{[+]}$ in order to overtake the slow ones. However, the
presence of heavy vehicles decreases the velocities of the overall
cars$^{[+]}$; leading therefore to a decrease of the lane-changing
rate.
\subsection{Rear-end collision}
A rear-end collision is a traffic accident where a vehicle hits
the rear of the vehicle in front of it. Rear-end collision is
mainly caused by inattentive driving behavior and not keeping
proper distance from the preceding vehicle. The resulting injuries
tended to be of low severity. In our model, the dangerous
situation corresponding to the rear-end collision exists when the
conditions 1 and 2 occur simultaneously. The probability of
rear-end collision (PRC) as a function of $\rho^{[+]}$, for
different values of $\rho^{[-]}$, is illustrated in figures 4a and
4b. First, we analyze the situation where $\rho^{[-]}$ is low
enough. When no heavy vehicle is present in the home lane, the PRC
is as like as in one-lane traffic model \cite{mou3}. That is,
there exists a critical density $\rho_{c}^{[+]}$ situated in the
low-density region, below which no rear-end collision can occur.
With increasing $\rho^{[+]}$, PRC increases, reaches a maximum,
then decreases and vanishes when $\rho^{[+]}$ exceeds some high
value $\rho_{h}^{[+]}$. Nevertheless, the presence of only one
heavy vehicle provokes an important increase of the risk of
rear-end collision over a wide range of the home lane density
(fig. 4a). Indeed, the presence of heavy vehicles provokes the
creation of traffic jams and platoons of cars in the home lane.
This will promotes the lane-changing of cars$^{[+]}$ in order to
overtake the leading heavy vehicles (see fig. 3a). In this
situation, rear-end collisions occur more likely, because a
vehicle suddenly decelerates due to an unexpected deceleration of
the jammed vehicles ahead.
\newline\ In figure 4b, we show simulation results for the PRC
in the case where $\rho^{[-]}$ is high enough. First, we analyze
the situation when the heavy vehicles are absent in the circuit.
So, we found that the risk of rear-end collision exists even at
very low home-lane density. If $\rho^{[+]}$ is low, free available
spaces exist on lane$^{[+]}$ and some cars$^{[-]}$ pass other
cars$^{[-]}$ and then impede traffic on lane$^{[+]}$. This causes
an important deceleration manoeuvres of vehicles$^{[+]}$. Thus, in
contrast to one-lane unidirectional models, an important value of
the risk of rear-end collision will be observed, even at very low
values of $\rho^{[+]}$. When heavy vehicles are present, the
fraction of both passing cars$^{[+]}$ and passing cars$^{[-]}$
will increase; leading therefore to the creation of more jams in
both lanes. This may enhance the probability of the occurrence of
rear-end dangerous situations.
\subsection{Lane-changing collision}
In addition to rear-end collisions, personal injuries often times
result when one vehicle makes an improper lane change and makes
contact with the vehicle in the adjacent lane. The lane-changing
collision produces injuries and damage which are often severe. In
our model, lane-changing collision occurs more likely when a
passing car$^{[+]}$ performs an unexpected return to the home
lane. The dangerous situation corresponding to the lane-changing
collision exists when the conditions 3 and 4 occur simultaneously.
Thus, it is obvious that the probability of lane-changing
collision (PLC) depends straightforwardly on the frequency of
lane-changing of cars$^{[+]}$. The PLC as a function of
$\rho^{[+]}$, for different values of $\rho^{[-]}$, is illustrated
in figures 5a and 5b. In figure 5a, we show simulation results for
the PLC in the case where $\rho^{[-]}$ is low enough. So, we
observe that, at low values of $\rho^{[+]}$, the risk of
lane-changing collision is very low in the case where heavy
vehicles are absent; but it is very high when at least one heavy
vehicle is present in the circuit. At very low densities, we found
that increasing the number of heavy vehicles has no effect on the
rate of lane-changing accidents. At relatively high density of
vehicles$^{[+]}$, the influence of the number of heavy vehicles is
noticeable. That is, we see that the PLC increases considerably
with increasing the fraction of heavy vehicles. At higher
densities, the influence of heavy vehicles becomes weak and
disappears completely beyond certain value of $\rho^{[+]}$.
\newline\ In figure 5b, we show simulation results for the PLC in the case
where $\rho^{[-]}$ is high enough. We first observe that the risk
of lane-changing collision is very low compared to the case where
$\rho^{[-]}$ is low enough. We find also that the PLC is almost
nonexistent when there is no heavy vehicles in the circuit. The
PLC increases with increasing the number of heavy vehicles; and
more importantly near the critical density.
\subsection{Head-on collision}
A head-on collision typically occurs when a vehicle executes a
passing manoeuvre and crashes into an oncoming vehicle. These type
of collisions are often fatal. The occurrence of both conditions 5
and 6 can be a potentially dangerous situation of head-on
collision. The probability of head-on collision (PHC) as a
function of the density $\rho^{[+]}$, for different values of
$\rho^{[-]}$, is illustrated in figures 6a and 6b. In figure 6a,
we show simulation results for low values of $\rho^{[-]}$. Let us
first describe the case where the heavy vehicles are absent. So,
when $\rho^{[+]}$ is below $\rho_{c}^{[+]}$, head-on collisions
occur most unlikely. Above $\rho_{c}^{[+]}$, the PHC increases
with increasing more the density, reaches a maximum, and then
decreases until it vanishes when $\rho^{[+]}$ exceeds some high
density value. When heavy vehicles are present in the circuit, a
peak appeared in the probability diagram, which is located in the
very low density region. Indeed, when both $\rho^{[-]}$ and
$\rho^{[+]}$ are low, the condition "Space2" occurs more likely if
the heavy vehicles are absent and more unlikely when at least one
heavy vehicle is present. With increasing more $\rho^{[+]}$, the
PHC decreases and vanishes when $\rho^{[+]}$ exceeds some high
density value. At fixed relatively high density, the PHC increases
with increasing the number of heavy vehicles present in the road.
Clearly, the increase of the number of heavy vehicles in the
circuit gives rise to the formation of additional jams behind
heavy vehicles. This encourages cars$^{[+]}$ to perform
lane-changing in order to pass the heavy vehicles. At the same
time, the occurrence of jams in the home lane will put at
disadvantage the return of passing cars$^{[+]}$. Therefore, at
relatively high $\rho^{[+]}$, the increase of the proportion of
heavy vehicles will increase the risk of head-on collision.
\newline\ In figure 6b, we show simulation results for the PHC when
$\rho^{[-]}$ is relatively high. We find that the PHC exhibits a
single peak around some low value of $\rho^{[+]}$. This means that
head-on collision occurs most likely if $\rho^{[+]}$ is very low.
Furthermore, we observe that the peak is more pronounced if no
heavy vehicle is present in the road. Indeed, at very low
densities, the frequency of lane-changing of cars$^{[+]}$ is more
important in the case of the presence of heavy vehicles than in
the case of their absence (Fig.3b). In these circumstances, the
presence of only one heavy vehicle in the circuit can reduce
significantly the risk of head-on collision. We find also that PHC
is insensitive to the variation of the proportion of heavy
vehicles. Indeed, when $\rho^{[-]}$ is relatively high, most of
head-on collisions are provoked by passing oncoming cars$^{[-]}$
which faced the vehicles$^{[+]}$ on the home lane. When the home
lane density becomes relatively high, no interaction will exist
between lanes and then no head-on collision will occur.
\section{Conclusion}
In summary, we have investigated the risk of collisions between
vehicles in two-lane bidirectional traffic flow models. We present
conditions for the occurrence of three different collisions:
rear-end collision, lane-changing collision and head-on collision.
The density of vehicles on the two lanes and the proportion of
heavy vehicles play an important role in the risk of car accident.
\newline\ When the density on the oncoming lane is low, more available free
space exist on the oncoming lane. If the home lane density is
relatively high, the fast cars are intended to overcome the slow
ones. This provokes the occurrence of dangerous situations in both
lanes. So, in these situations, head-on collisions between passing
cars and oncoming cars are very likely to occur. The unexpected
return of cars to their home lane may cause lane-changing
collisions. The rear-end collision may also occur because of the
formation of additional jams on the home lane. It is found that,
in bi-directional traffic, the probability of the rear-end
collision is significantly higher than that of the head-on
collision or the lane-changing collision. Nevertheless, the two
last collisions are an often fatal type of road traffic accident
whereas the rear-end collision may not result in catastrophic
injury. Slowly moving heavy vehicles cause additional delay on the
home lane and thus prevent the fluidity of traffic flow. This
provokes a diminution of the cars speeds, and then promotes the
pass of cars on the oncoming lane. Therefore, the risk of vehicle
collisions should increase with increasing the proportion of heavy
vehicles.
\newline\ When the density of the oncoming lane is relatively
high, oncoming cars can pass on the home lane whenever the density
of this last is low. This can provoke collisions with vehicles
travelling on the home lane. Here also, the presence of heavy
vehicle can enhance the risk of car accidents; especially for the
rear-end and the head-on collisions.
\newline\ Because it is sometimes impossible to evaluate
the risk of collisions using real crash data, the present study
provides a simulation approach to investigate the probability of
collision for bidirectional traffic accidents.
\newpage\

\newpage\ \textbf{Figures captions}

\begin{quote}
\textbf{Fig.1}. Illustration of the quantities relevant for the
lane changing rules in the two-lane bidirectional traffic system.

\textbf{Fig.2}. Flow of vehicles$^{[+]}$ as a function of the
density of vehicles$^{[+]}$: (a) The density of cars$^{[-]}$ is
low enough ($\rho^{[-]}=0.05$), (b) The density of cars$^{[-]}$ is
high enough ($\rho^{[-]}=0.30$).

\textbf{Fig.3}. The dependence on the density of vehicles$^{[+]}$
of the lane-changing frequency of cars$^{[+]}$: (a) The density of
cars$^{[-]}$ is low enough ($\rho^{[-]}=0.05$), (b) The density of
cars$^{[-]}$ is high enough ($\rho^{[-]}=0.30$).

\textbf{Fig.4}. The dependence on the density of vehicles$^{[+]}$
of the probability of rear-end collision: (a) The density of
cars$^{[-]}$ is low enough ($\rho^{[-]}=0.05$), (b) The density of
cars$^{[-]}$ is high enough ($\rho^{[-]}=0.30$).

\textbf{Fig.5.} The dependence on the density of vehicles$^{[+]}$
of the probability of lane-changing collision: (a) The density of
cars$^{[-]}$ is low enough ($\rho^{[-]}=0.05$), (b) The density of
cars$^{[-]}$ is high enough ($\rho^{[-]}=0.30$).

\textbf{Fig.6}. The dependence on the density of vehicles$^{[+]}$
of the probability of head-on collision: (a) The density of
cars$^{[-]}$ is low enough ($\rho^{[-]}=0.05$), (b) The density of
cars$^{[-]}$ is high enough ($\rho^{[-]}=0.30$).

\end{quote}

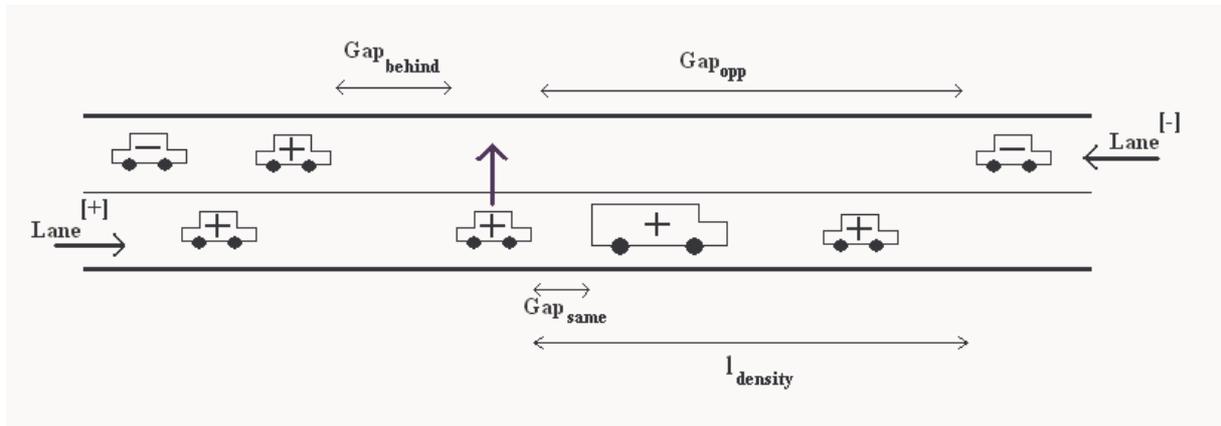
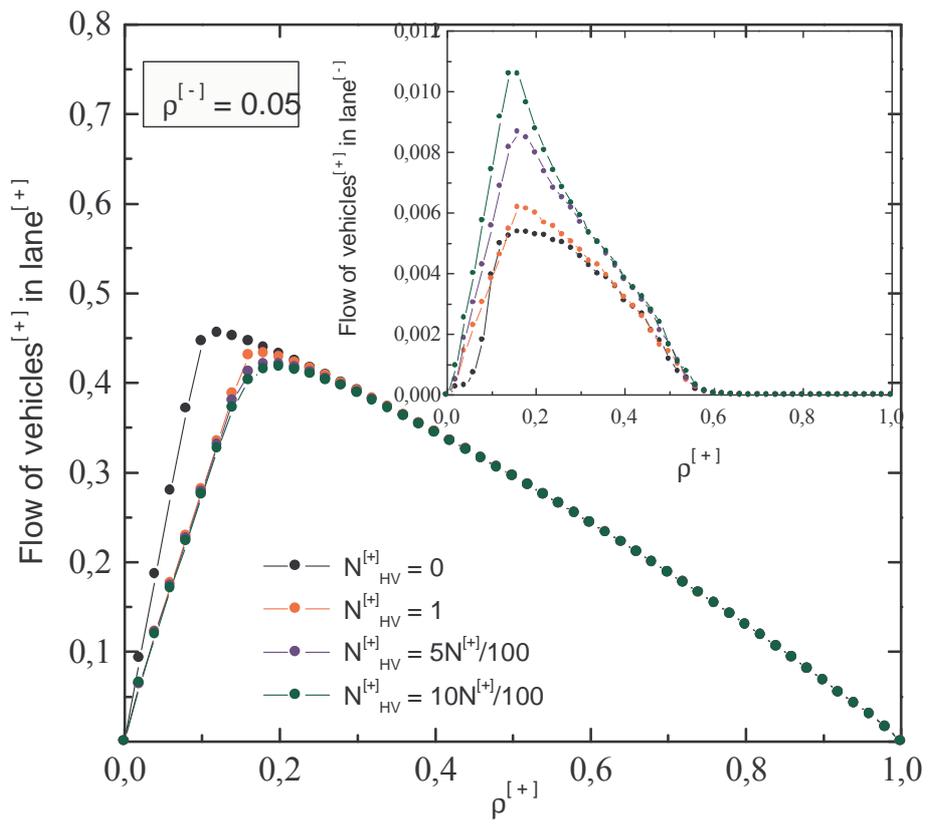
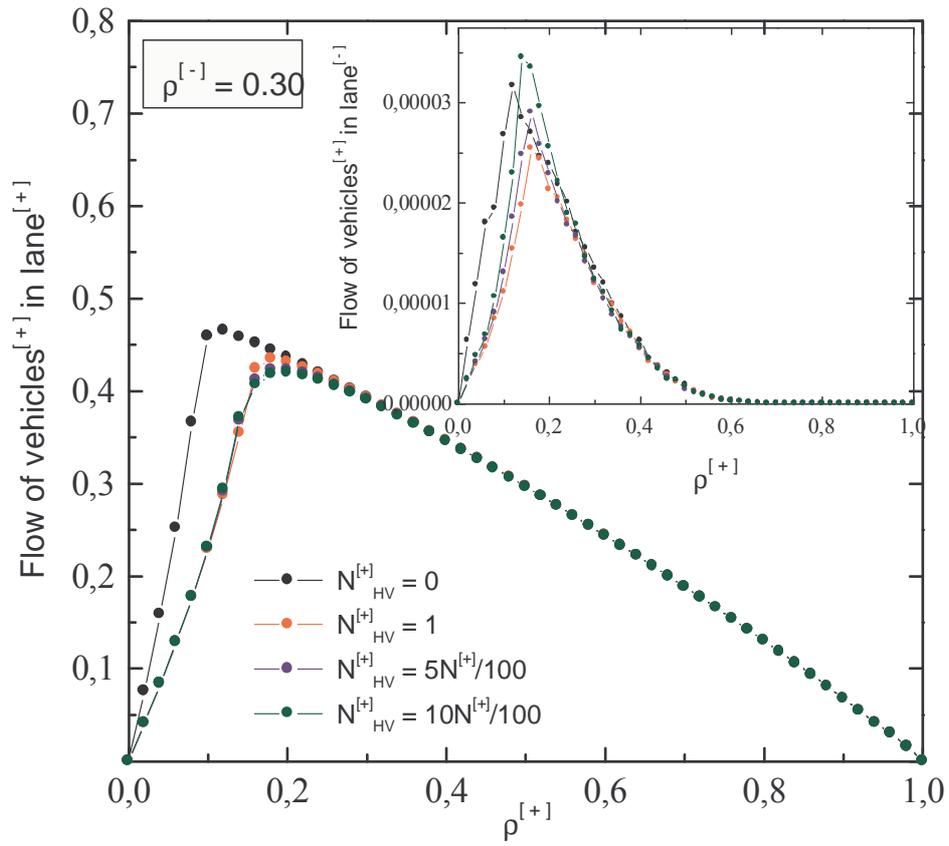
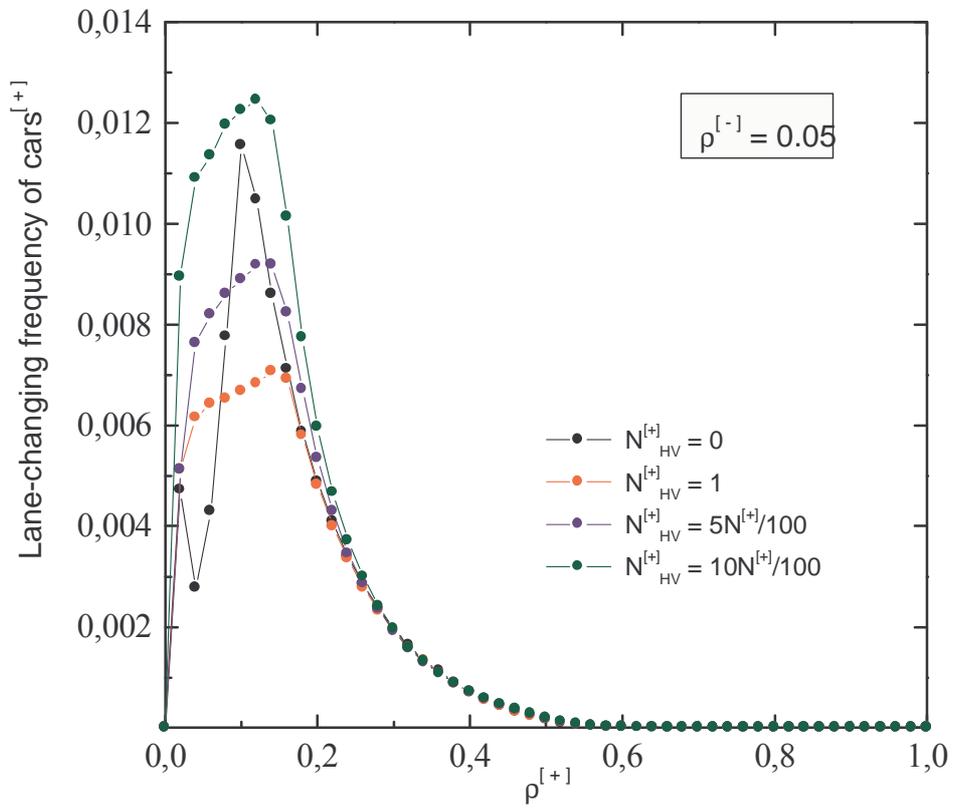
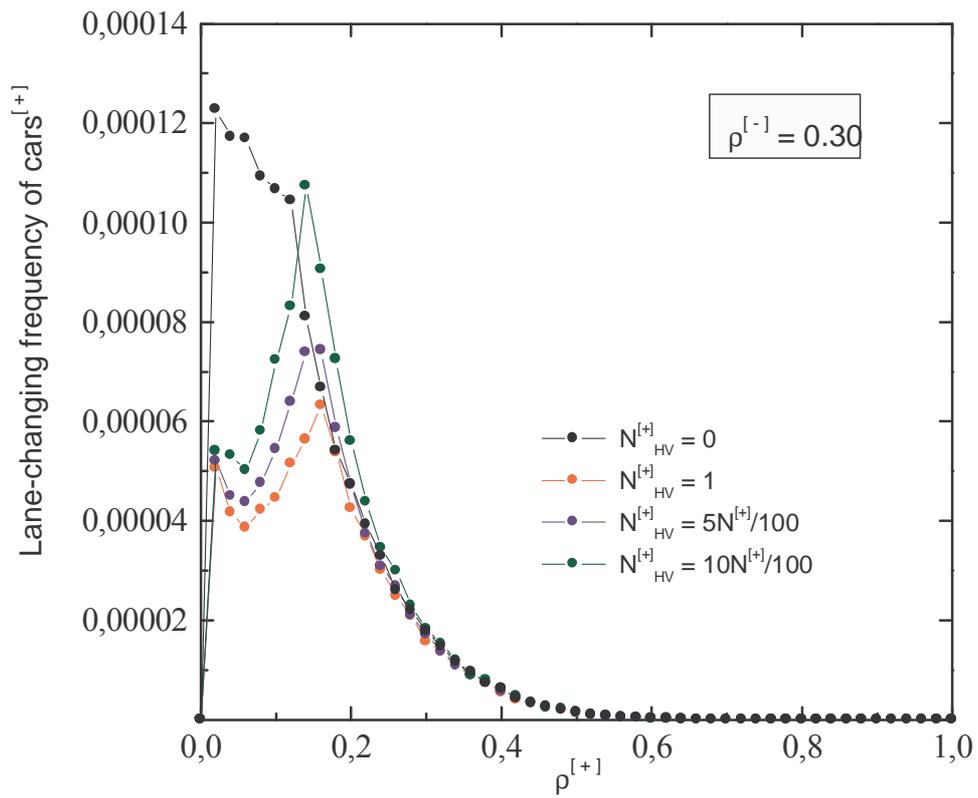
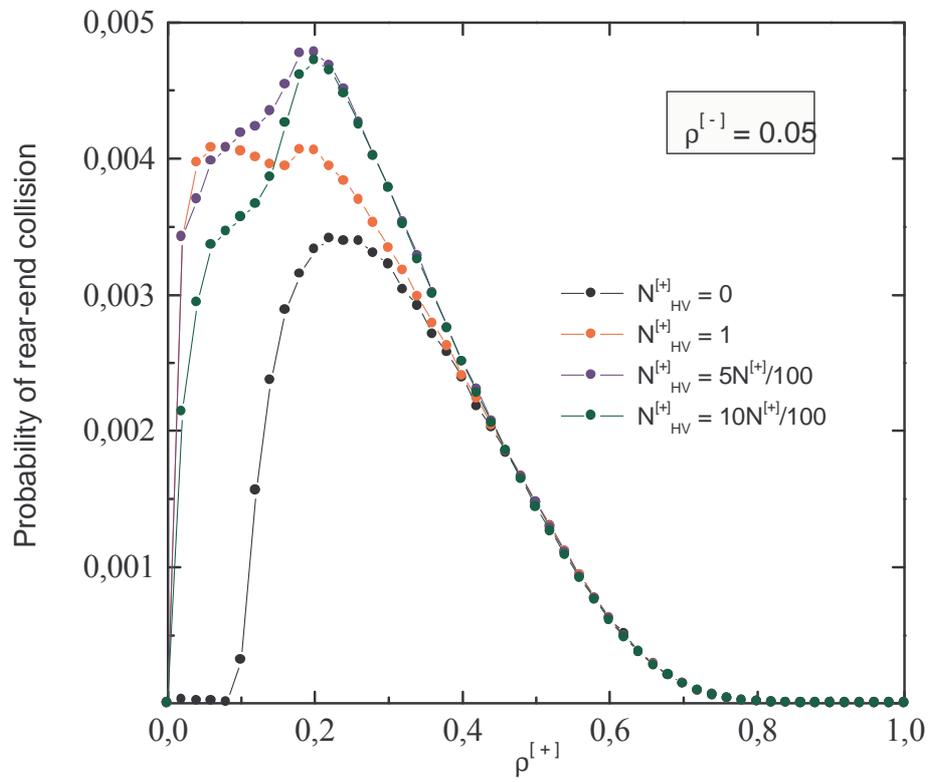
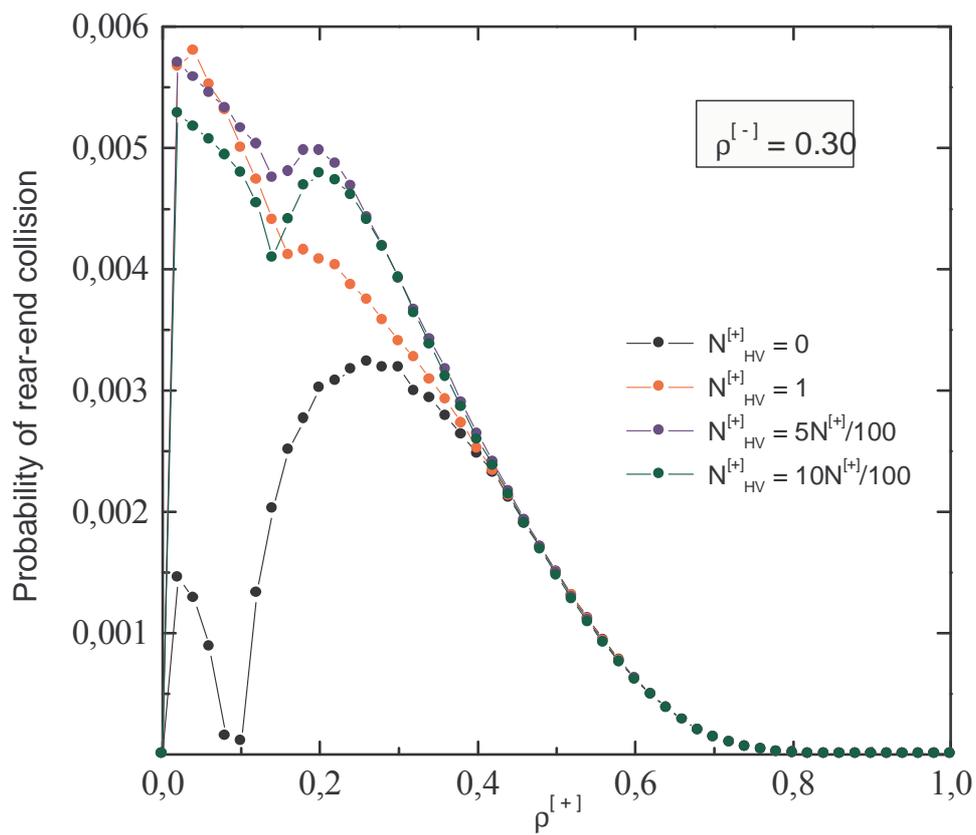
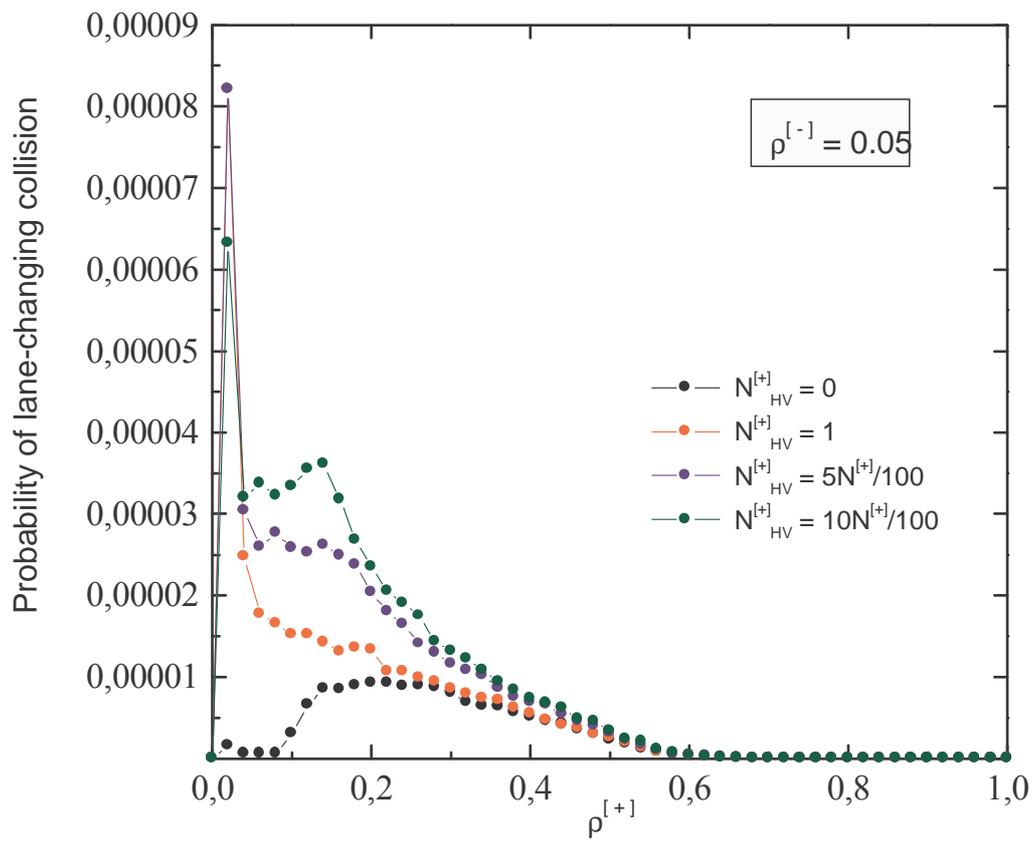
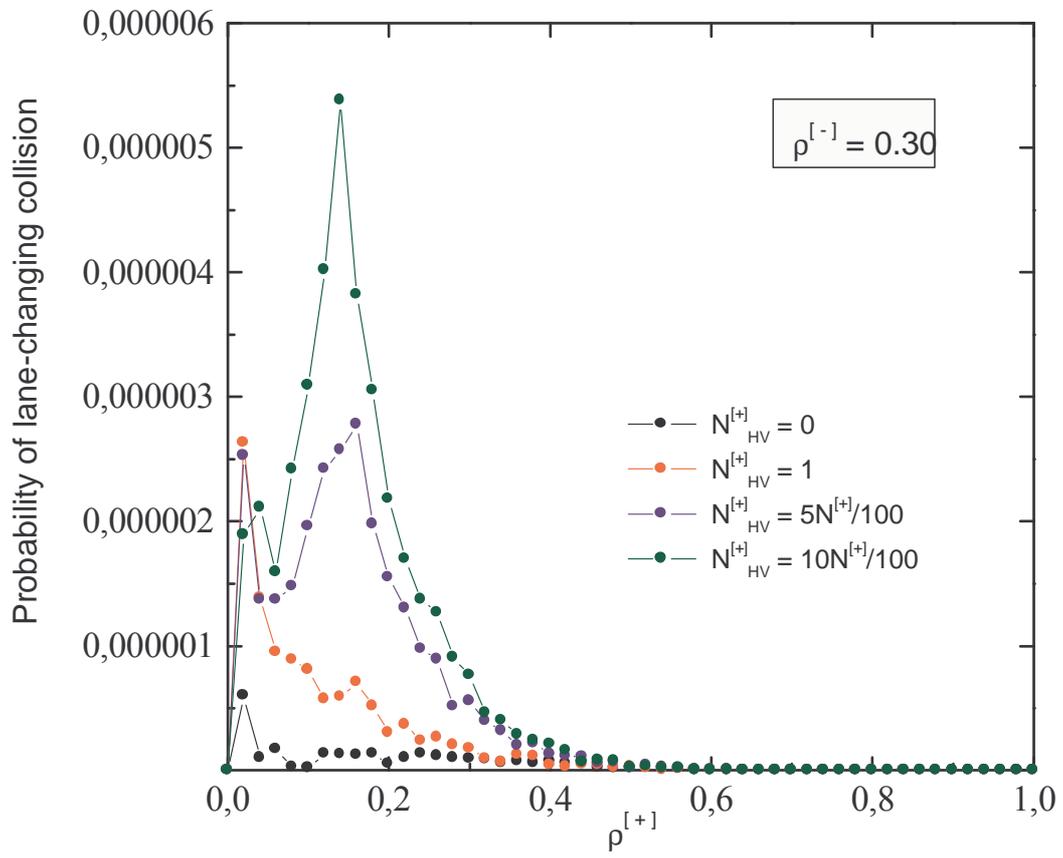
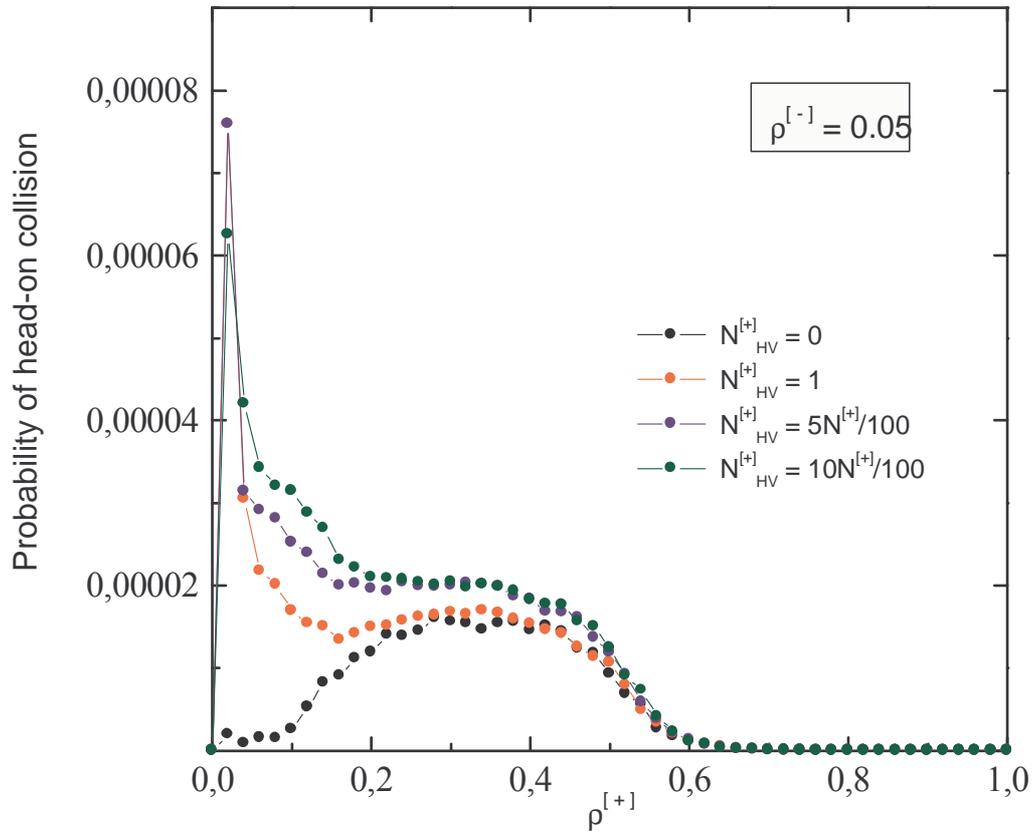
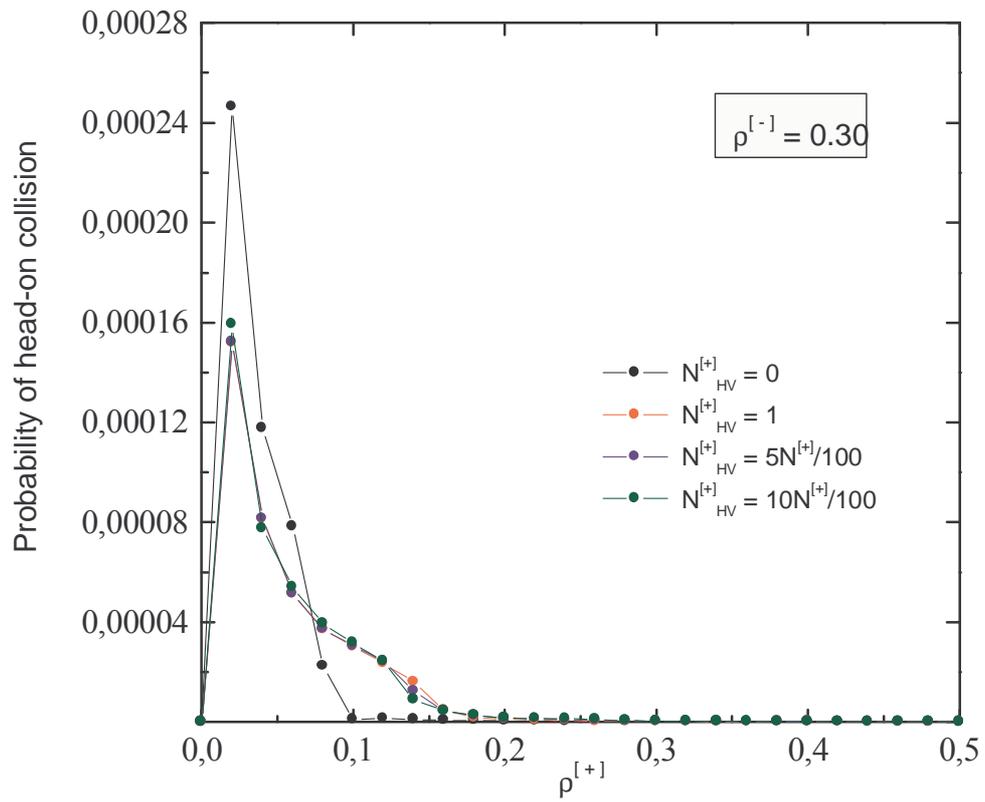
\end{document}